\documentclass[twocolumn,showpacs,aps,prd]{revtex4}

\usepackage[dvips]{graphicx}
\usepackage{epsf}  % for embedding postscript figures

\usepackage{graphicx}
\usepackage{dcolumn}
\usepackage{epsfig}
\usepackage{psfrag}
\usepackage{amsmath}

\newcommand{\psfile}[3][]{ % (arguments: [BoundingBox] full filename, 
  \begin{center}
    \setlength{\epsfxsize}{#3\linewidth}\leavevmode
    \def\noOpt{}\def\testit{#1}\ifx\testit\noOpt%
      \epsfbox{#2}%
    \else%
      \epsfbox[#1]{#2}%
    \fi
  \end{center} 
}

\newcommand{\BaBarYear}    {11}
\newcommand{\BaBarNumber}  {010}

\newcommand{\SLACPubNumber} {14842}

 \newcommand{\BaBarType}      {PUB}  % Journal publication

\input pubboard/babarsym 

\providecommand{\calB}{\ensuremath{{\cal B}}}
\newcommand{\calH}{\ensuremath{{\cal H}}}
\newcommand{\calO}{\ensuremath{{\cal O}}}
\providecommand{\calP}{\ensuremath{{\cal P}}}

\newcommand{\bfemsix}{${\cal B}(10^{-6}$)}

\newcommand{\msp}{\phantom{1}}

\newcommand{\bfig}{\begin{figure}[htbpc!]}
\newcommand{\efig}{\end{figure}}

\def\dbline{\noalign{\vskip 0.10truecm\hrule}\noalign{\vskip 2pt}\noalign{\hrule\vskip 0.10truecm}}
\def\sgline{\noalign{\vskip 0.10truecm\hrule\vskip 0.10truecm}}

\def\etal{{\it et al.}}

\providecommand{\etapr}{\ensuremath{\eta^\prime}}

\newcommand{\DE}{\ensuremath{\Delta E}}

\newcommand{\xf}{\ensuremath{{\cal F}}}

\newcommand{\costhr}{\ensuremath{\cos\theta_{\rm T}}}
\providecommand{\KS}{\ensuremath{K_S^0}}

\providecommand{\UfourS}{\ensuremath{\Upsilon(4S)}}
\providecommand{\pvec}{{\bf p}}
\providecommand{\half}{\ensuremath{{1\over2}}}

\newcommand{\rhoz}{\ensuremath{\rho^0}}
\newcommand{\rhop}{\ensuremath{\rho^+}}

\def\qqbar{\ensuremath{q\bar q}}

\def\beq{\begin{equation}}
\def\eeq{\end{equation}}
\def\bef{\begin{figure}}
\def\edf{\end{figure}}
\def\ben{\begin{enumerate}}
\def\een{\end{enumerate}}
\def\bear{\begin{array}}
\def\enar{\end{array}}
\def\beqa{\begin{eqnarray}}
\def\eeqa{\end{eqnarray}}
\def\bit{\begin{itemize}}
\def\eit{\end{itemize}}
\def\to{\ensuremath{\rightarrow}}

\def\qq{\ensuremath{q\overline q}}

\def\DE{\ensuremath{{\Delta E}}}

\def\mes{\ensuremath{{m_{ES}}}}

\newcommand{\acp}{\ensuremath{\calA_{ch}}}
\newcommand{\aonem}{\ensuremath{a_1^{-}}\xspace}
\newcommand{\rhom}{\ensuremath{\rho^{-}}\xspace}
\newcommand{\fz}{\ensuremath{f_{0}}\xspace}
\newcommand{\ftwo}{\ensuremath{f_{2}(1270)}\xspace}
\newcommand{\Kst}{\ensuremath{K^{*}}\xspace}
\newcommand{\Kstz}{\ensuremath{K^{*0}}\xspace}

\newcommand{\Kstp}{\ensuremath{K^{*+}}\xspace}

\newcommand{\KstZeroRes}{\ensuremath{K_0^{*}(1430)}\xspace}
\newcommand{\KstZero}{\ensuremath{(K\pi)_0^*}\xspace}

\newcommand{\KstzZeroRes}{\ensuremath{K_0^{*}(1430)^0}\xspace}

\newcommand{\KstzZero}{\ensuremath{(K\pi)_0^{*0}}\xspace}
\newcommand{\KstpZero}{\ensuremath{(K\pi)_0^{*+}}\xspace}
\newcommand{\KstOne}{\ensuremath{K^{*}(892)}\xspace}

\newcommand{\KstzOne}{\ensuremath{K^{*}(892)^0}\xspace}

\newcommand{\KstpOne}{\ensuremath{K^{*}(892)^+}\xspace}
\newcommand{\KstTwo}{\ensuremath{K_2^{*}(1430)}\xspace}

\newcommand{\KstzTwo}{\ensuremath{K_2^{*}(1430)^0}\xspace}

\newcommand{\frhozKstz}{\ensuremath{\rho^0 K^{*0}}\xspace}
\newcommand{\rhozKstz}{\ensuremath{B^0\ra\frhozKstz}\xspace}
\newcommand{\frhozKstzz}{\ensuremath{\rho^0 K_0^{*0}}\xspace}
\newcommand{\frhozKstzZeroRes}{\ensuremath{\rho^0 K_0^{*}(1430)^0}\xspace}
\newcommand{\frhozKstzZero}{\ensuremath{\rho^0 \KstzZero}\xspace}
\newcommand{\rhozKstzZero}{\ensuremath{\Bz\ra\frhozKstzZero}\xspace}
\newcommand{\rhozKstzZeroRes}{\ensuremath{\Bz\ra\frhozKstzZeroRes}\xspace}
\newcommand{\frhozKstzOne}{\ensuremath{\rho^0 K^{*}(892)^0}\xspace}
\newcommand{\rhozKstzOne}{\ensuremath{\Bz\ra\frhozKstzOne}\xspace}
\newcommand{\frhozKstzTwo}{\ensuremath{\rho^0 K_2^{*}(1430)^0}\xspace}

\newcommand{\ffzKstz}{\ensuremath{f_0 K^{*0}}\xspace}
\newcommand{\fzKstz}{\ensuremath{B^0\ra\ffzKstz}\xspace}
\newcommand{\ffzKstzz}{\ensuremath{f_0 K_0^{*0}}\xspace}
\newcommand{\ffzKstzZeroRes}{\ensuremath{f_0 K_0^{*}(1430)^0}\xspace}
\newcommand{\ffzKstzZero}{\ensuremath{f_0 \KstzZero}\xspace}
\newcommand{\fzKstzZero}{\ensuremath{\Bz\ra\ffzKstzZero}\xspace}
\newcommand{\fzKstzZeroRes}{\ensuremath{\Bz\ra\ffzKstzZeroRes}\xspace}
\newcommand{\ffzKstzOne}{\ensuremath{f_0 K^{*}(892)^0}\xspace}
\newcommand{\fzKstzOne}{\ensuremath{\Bz\ra\ffzKstzOne}\xspace}
\newcommand{\ffzKstzt}{\ensuremath{f_0 K_2^{*0}}\xspace}
\newcommand{\ffzKstzTwo}{\ensuremath{f_0 K_2^{*}(1430)^0}\xspace}
\newcommand{\fzKstzTwo}{\ensuremath{B^0\ra\ffzKstzTwo}\xspace}
\newcommand{\fftwoKstzOne}{\ensuremath{f_2(1270) K^*(892)^0}\xspace}

\newcommand{\ftwoKstzOne}{\ensuremath{\Bz\ra\fftwoKstzOne}\xspace}
\newcommand{\frhomKstp}{\ensuremath{\rho^- K^{*+}}\xspace}
\newcommand{\rhomKstp}{\ensuremath{B^0\ra\frhomKstp}\xspace}
\newcommand{\frhomKstpz}{\ensuremath{\rho^- K_0^{*+}}\xspace}
\newcommand{\frhomKstpZeroRes}{\ensuremath{\rho^- K_0^{*}(1430)^+}\xspace}
\newcommand{\frhomKstpZero}{\ensuremath{\rho^- \KstpZero}\xspace}
\newcommand{\rhomKstpZero}{\ensuremath{\Bz\ra\frhomKstpZero}\xspace}
\newcommand{\rhomKstpZeroRes}{\ensuremath{\Bz\ra\frhomKstpZeroRes}\xspace}

\newcommand{\frhomKstpOne}{\ensuremath{\rho^- K^{*}(892)^+}\xspace}
\newcommand{\rhomKstpOne}{\ensuremath{\Bz\ra\frhomKstpOne}\xspace}

\newcommand{\frhomKstpTwo}{\ensuremath{\rho^- K_2^{*}(1430)^+}\xspace}

\newcommand{\faonemKp}{\ensuremath{a_1^- K^{+}}\xspace}

\newcommand{\aonemKp}{\ensuremath{\Bz\ra\faonemKp}\xspace}

\renewcommand{\pipi}{\ensuremath{\pi\pi}\xspace}

\newcommand{\fL}{\ensuremath{f_{L}}\xspace}

\newcommand{\nbb}        {\ensuremath{(471.0\pm 2.8)\times 10^{6}}\xspace}

\newcommand{\mkpi}{\ensuremath{m_{K\pi}}\xspace}

\newcommand{\mpipi}{\ensuremath{m_{\pi\pi}}\xspace}
\newcommand{\mpipiC}{\ensuremath{m_{\pip\pim}}\xspace}

\newcommand{\sigfz}{\ensuremath{\sigma/f_0(600)}\xspace}

\newcommand{\BrhozKstz}{\ensuremath{\calB(\rhozKstz)}}
\newcommand{\nrhozKstz}{\ensuremath{376\pm 37}} 	% signal yield
\newcommand{\biasrhozKstz}{\ensuremath{44\pm3}} 	% bias on yield
\newcommand{\rrhozKstz}{\ensuremath{5.1\pm 0.6 ^{+0.6}_{-0.8} }} % BF measurement
\newcommand{\srhozKstz}{\ensuremath{6.0}}             		% significance
\newcommand{\ArhozKstz}{\ensuremath{\msp-0.06\pm 0.09\pm 0.02}}	% Ach
\newcommand{\fLrhozKstz}{\ensuremath{\msp0.40\pm 0.08\pm 0.11}}	% fL
\newcommand{\RrhozKstz}{\ensuremath{(\rrhozKstz)\times 10^{-6}}}
\newcommand{\BfzKstz}{\ensuremath{\calB(\fzKstz)\times\calB(\fz\ra\pi\pi)}}
\newcommand{\rfzKstz}{\ensuremath{5.7\pm 0.6 \pm 0.4}} % BF measurement
\newcommand{\nfzKstz}{\ensuremath{220\pm 23}} 	% signal yield
\newcommand{\biasfzKstz}{\ensuremath{2.1\pm1.6}} 	% bias
\newcommand{\sfzKstz}{\ensuremath{9.8}}             		% significance
\newcommand{\AfzKstz}{\ensuremath{\msp+0.07\pm 0.10\pm 0.02}}	% Ach
\newcommand{\RfzKstz}{\ensuremath{(\rfzKstz)\times 10^{-6}}}

\newcommand{\BrhomKstp}{\ensuremath{\calB(\rhomKstp)}}
\newcommand{\rrhomKstp}{\ensuremath{10.3\pm 2.3\pm 1.3}} % BF measurement
\newcommand{\nrhomKstp}{\ensuremath{167\pm 27}} 	% signal yield
\newcommand{\biasrhomKstp}{\ensuremath{23\pm 3}} 	% bias
\newcommand{\srhomKstp}{\ensuremath{5.1}}             		% significance
\newcommand{\ArhomKstp}{\ensuremath{\msp+0.21\pm 0.15\pm 0.02}}  % Ach
\newcommand{\fLrhomKstp}{\ensuremath{\msp0.38\pm0.13\pm 0.03}}	% fL
\newcommand{\RrhomKstp}{\ensuremath{(\rrhomKstp)\times 10^{-6}}}

\newcommand{\BrhozKstzZero}{\ensuremath{\calB(\rhozKstzZero)\times\calB(\KstZero\ra K\pi)}}
\newcommand{\rrhozKstzZero}{\ensuremath{31\pm 4\pm 3}\xspace}     % BF (e-6)
\newcommand{\nrhozKstzZero}{\ensuremath{1045\pm 36\pm 118}\xspace}  % yield in high mass sideband
\newcommand{\biasrhozKstzZero}{\ensuremath{80\pm 11}\xspace}  % bias
\newcommand{\srhozKstzZero}{\ensuremath{6.3}}            % significance
\newcommand{\RrhozKstzZero}{\ensuremath{(\rrhozKstzZero)\times 10^{-6}}}
\newcommand{\BrhozKstzZeroRes}{\ensuremath{\calB(\rhozKstzZeroRes)}}
\newcommand{\rrhozKstzZeroRes}{\ensuremath{27\pm 4\pm 2\pm 3}\xspace}     % BF (e-6)
\newcommand{\RrhozKstzZeroRes}{\ensuremath{(\rrhozKstzZeroRes)\times 10^{-6}}}
\newcommand{\nfzKstzZero}{\ensuremath{88\pm 19\pm 10}\xspace}
\newcommand{\biasfzKstzZero}{\ensuremath{7\pm 1}\xspace}
\newcommand{\rfzKstzZero}{\ensuremath{3.1\pm 0.8\pm 0.7}\xspace}
\newcommand{\sfzKstzZero}{\ensuremath{3.0}}             % significance
\newcommand{\RfzKstzZero}{\ensuremath{(\rfzKstzZero)\times 10^{-6}}}
\newcommand{\BfzKstzZeroRes}{\ensuremath{\calB(\fzKstzZeroRes)\times\calB(\fz\ra\pi\pi)}}
\newcommand{\rfzKstzZeroRes}{\ensuremath{2.7\pm 0.7\pm 0.5\pm 0.3}\xspace}
\newcommand{\RfzKstzZeroRes}{\ensuremath{(\rfzKstzZeroRes)\times 10^{-6}}}
\newcommand{\BfzKstzTwo}{\ensuremath{\calB(\fzKstzTwo)\times\calB(\fz\ra\pi\pi)}}
\newcommand{\nfzKstzTwo}{\ensuremath{134\pm 14\pm 23}\xspace}
\newcommand{\biasfzKstzTwo}{\ensuremath{0\pm 2}\xspace}
\newcommand{\rfzKstzTwo}{\ensuremath{8.6\pm 1.7\pm 1.0}\xspace}
\newcommand{\sfzKstzTwo}{\ensuremath{4.3}}             	% significance
\newcommand{\RfzKstzTwo}{\ensuremath{(\rfzKstzTwo)\times 10^{-6}}}
\newcommand{\BrhomKstpZero}{\ensuremath{\calB(\rhomKstpZero)\times\calB(\KstZero\ra K\pi)}}
\newcommand{\nrhomKstpZero}{\ensuremath{221\pm 74}\xspace}  %signal yield
\newcommand{\biasrhomKstpZero}{\ensuremath{-5\pm 8}\xspace} %fit bias
\newcommand{\rrhomKstpZero}{\ensuremath{32\pm 10\pm 6}\xspace}
\newcommand{\srhomKstpZero}{\ensuremath{2.8}}           % significance
\newcommand{\ulrhomKstpZero}{\ensuremath{48}}         % 90% CL upper limit
\newcommand{\RrhomKstpZero}{\ensuremath{(\rrhomKstpZero)\times 10^{-6}}}
\newcommand{\BrhomKstpZeroRes}{\ensuremath{\calB(\rhomKstpZeroRes)}}
\newcommand{\rrhomKstpZeroRes}{\ensuremath{28\pm 10\pm 5\pm 3}\xspace}
\newcommand{\RrhomKstpZeroRes}{\ensuremath{(\rrhomKstpZeroRes)\times 10^{-6}}}
\newcommand{\nftwoKstz}{\ensuremath{627\pm 41}\xspace}

\long\def\inst#1{\par\nobreak\kern 4pt\nobreak
    {\it #1}\par\vskip 10pt plus 3pt minus 3pt}

\newcommand{\theTitle}{{\boldmath \Bz meson decays to $\rhoz\Kstz$, $\fz\Kstz$, and $\rhom\Kstp$, including higher \Kst\ resonances}}

\begin{document}

\begin{flushleft}
\babar-\BaBarType-\BaBarYear/\BaBarNumber \\
SLAC-PUB-\SLACPubNumber \\
\end{flushleft}

\title{\theTitle}

%% author list as of 02-Apr-2011 (386 authors)
%
\author{J.~P.~Lees}
\author{V.~Poireau}
\author{V.~Tisserand}
\affiliation{Laboratoire d'Annecy-le-Vieux de Physique des Particules (LAPP), Universit\'e de Savoie, CNRS/IN2P3,  F-74941 Annecy-Le-Vieux, France}
\author{J.~Garra~Tico}
\author{E.~Grauges}
\affiliation{Universitat de Barcelona, Facultat de Fisica, Departament ECM, E-08028 Barcelona, Spain }
\author{M.~Martinelli$^{ab}$}
\author{D.~A.~Milanes$^{a}$}
\author{A.~Palano$^{ab}$ }
\author{M.~Pappagallo$^{ab}$ }
\affiliation{INFN Sezione di Bari$^{a}$; Dipartimento di Fisica, Universit\`a di Bari$^{b}$, I-70126 Bari, Italy }
\author{G.~Eigen}
\author{B.~Stugu}
\author{L.~Sun}
\affiliation{University of Bergen, Institute of Physics, N-5007 Bergen, Norway }
% ...
\collaboration{The \babar\ Collaboration}
\noaffiliation

\date{December 14, 2011}

\begin{abstract}
We present branching fraction measurements for the decays \rhozKstz, \fzKstz, and \rhomKstp, where \Kst\ is an $S$-wave \KstZero or a \KstOne meson; we also measure \fzKstzTwo.  For the \KstOne channels, we report measurements of longitudinal polarization fractions (for $\rho$ final states) and direct \CP-violation asymmetries.  These results are obtained from a sample of \nbb\ \BB\ pairs collected with the \babar\ detector at the PEP-II asymmetric-energy \epem\ collider at the SLAC National Accelerator Laboratory.  We observe \frhozKstzOne, \frhozKstzZero, \ffzKstzOne, and \frhomKstpOne with greater than $5\sigma$ significance, including systematics.  We report first evidence for \ffzKstzZero and \ffzKstzTwo, and place an upper limit on \frhomKstpZero.  Our results in the \KstOne channels are consistent with no direct \CP\-violation.
\end{abstract}


\maketitle
\newpage

\setcounter{footnote}{0}

%-----------------------------------------
\section{Introduction}\label{sec:intro}
%-----------------------------------------

Measurements of the branching fractions and angular distributions of $B$ meson decays to hadronic final states without a charm quark probe the dynamics of both the weak and strong interactions.  Such studies also play an important role in understanding \CP\ violation in the quark sector and in searching for evidence for physics beyond the Standard Model~\cite{cheng_smith}. 

We report measurements of branching fractions for the decays \rhozKstzOne, \fzKstzOne, \rhomKstpOne, \rhozKstzZero, \fzKstzZero, \rhomKstpZero, and \fzKstzTwo.  For the $\rho\KstOne$ channels we measure the longitudinal fraction \fL, and for all \KstOne\ channels we measure charge asymmetries \acp.  The notation $\rho$ refers to the $\rho(770)$~\cite{PDG} and \fz to the $\fz(980)$~\cite{fzero}.  Throughout this paper we use \Kst\ to refer to any of the scalar \KstZero, vector \KstOne, or tensor \KstTwo states~\cite{PDG}.  The notation \KstZero refers to the scalar $K\pi$, which we describe with a LASS model~\cite{LASS,Latham}, combining the \KstZeroRes resonance with an effective-range non-resonant component.  Charge-conjugate modes are implied throughout this paper.  

\begin{figure}[!htbp]
\begin{center}
  \includegraphics[width=1.0\linewidth]{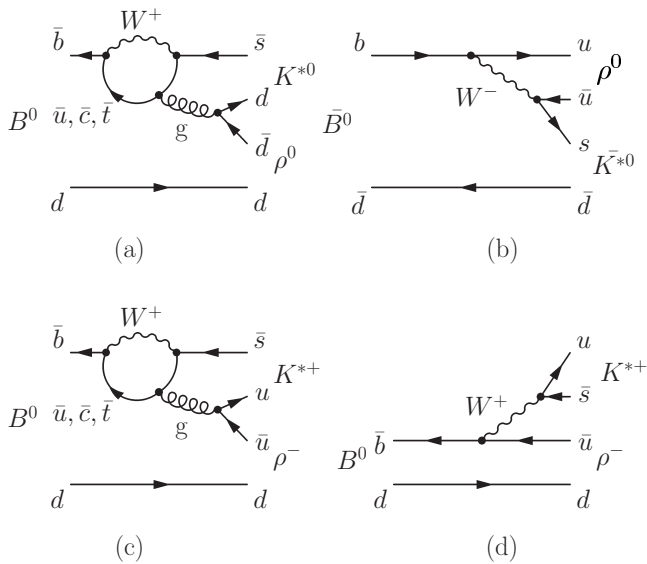}
  \caption{
   Feynman diagrams for (a--b) \rhozKstz and (c--d) \rhomKstp.  Gluonic penguin diagrams (a, c) dominate over tree (b, d) contributions.
   }
  \label{fig:feyn}
\end{center}
\end{figure}

The charmless decays $\B\ra\rho\Kst$ proceed through dominant penguin loops and CKM-suppressed tree processes ($\Bp\ra\rhop\Kstz$ is pure penguin), as shown in Fig.~\ref{fig:feyn}.  Na\"ive factorization models predict a large longitudinal polarization fraction $f_L$ (of order $(1-4m_V^2/m_B^2)\sim 0.9$) for vector--vector ($VV$) decays, where $m_V$ and $m_B$ are the masses of the vector and $B$ mesons, respectively~\cite{cheng_smith}. However, measurements of penguin-dominated $VV$ decays, such as the previous measurements of \rhozKstzOne and $\Bp\ra\rhop\KstzOne$~\cite{babar_rhoKst2006,BelleRhoKst}, find $\fL\sim 0.5$; a recent \babar\ measurement of $\Bp\ra\rhoz\KstpOne$ finds $\fL=0.78\pm0.12$~\cite{FergusPRD}.  Recent predictions in QCD Factorization (QCDF)~\cite{QCDF} can accommodate $\fL\sim 0.5$, although correctly predicting both the branching fraction and \fL remains a challenge.  

Both the \babar\ and Belle Collaborations have previously measured the branching fractions of \rhozKstzOne and \fzKstzOne.  \babar\ has also placed a 90\% confidence level (C.L.) upper limit on \rhomKstpOne~\cite{babar_rhoKst2006,BelleRhoKst}.  Belle searched for non-resonant $\Bz\ra\rhoz\Kp\pim$ and $\fz\Kp\pim$ decays, finding a five standard deviation (5$\sigma$) significant result for $\rhoz\Kp\pim$~\cite{BelleRhoKst}.  Decays involving a $\rho$ or \fz along with a \KstZero or \KstTwo have not been the subject of previous studies.  Predictions exist from both QCDF~\cite{QCDF} and perturbative QCD (pQCD)~\cite{pQCD_Kst0} for the branching fractions (\calB) of the $\rho\KstZeroRes$ channels, with QCDF predicting values $\calO(\mbox{few}\times10^{-5})$ and pQCD $\calO(5\times10^{-7}-10^{-5})$.  Improved experimental measurements will help refine predictions and constrain physics beyond the Standard Model.

\begin{figure}[htbp]
\begin{center}
\scalebox{1.0}[1.0]{\includegraphics[angle=0,width=0.9\linewidth]{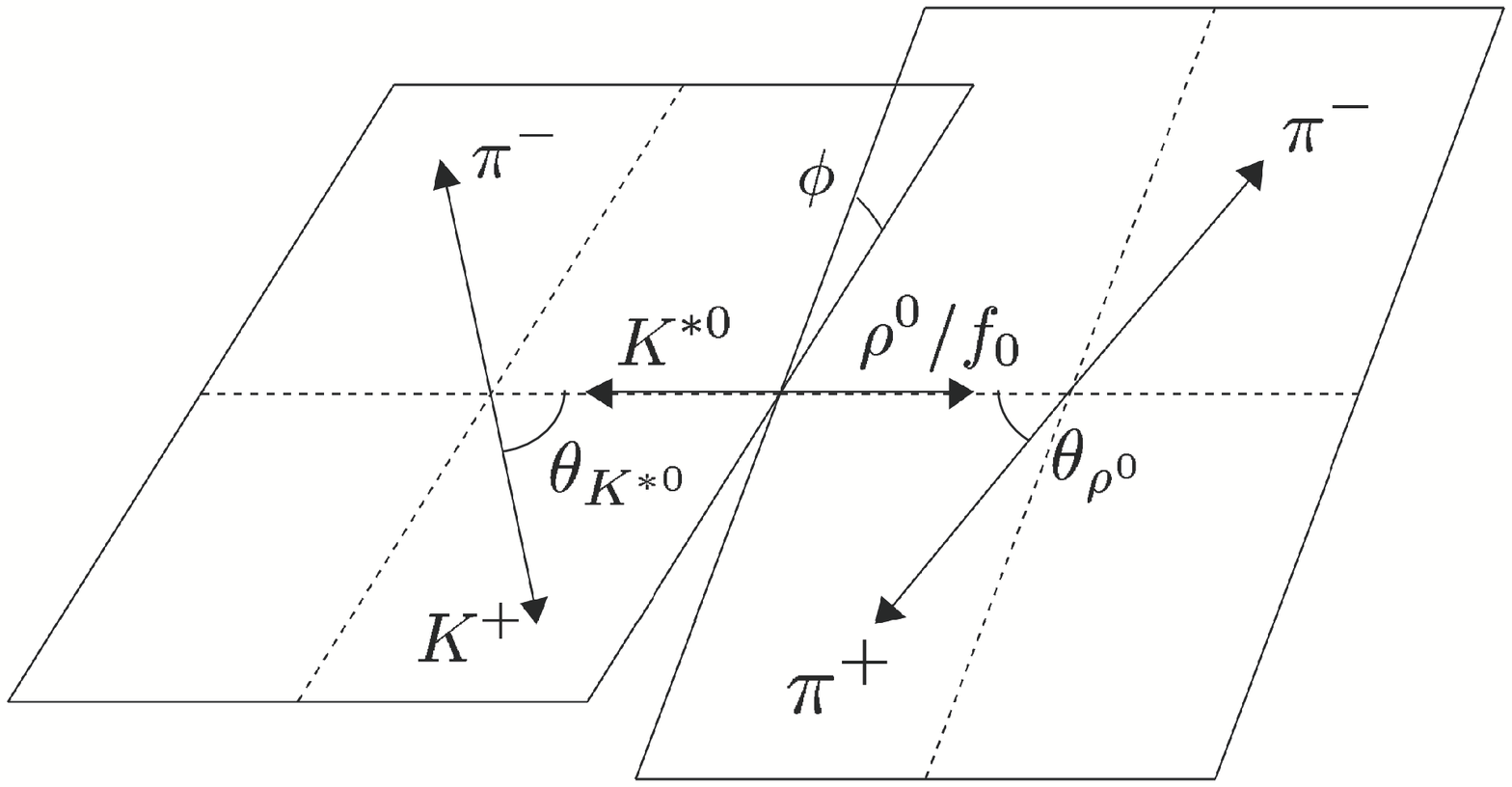}}
\end{center}
\vskip -0.5cm
\caption{\label{fig:helicity_angles}%
Definition of the helicity angles for \rhozKstz.
}
\end{figure}

The decays \rhozKstzOne and \rhomKstpOne are of the form $B\ra VV$; these decays have three polarization states, which are, in principle, accessible experimentally.  In practice, a full angular analysis requires a large number of signal events.  In the analyses described in this paper, we integrate over the azimuthal angle (the angle between the two vector meson decay planes).  The azimuthal angle is not correlated with any specific direction in the detector, so we assume a uniform acceptance over this angle.  We define the helicity angles $\theta_{\Kst}$ and $\theta_{\rho}$ and the azimuthal angle $\phi$ as shown in Fig.~\ref{fig:helicity_angles}.  
The helicity angles are defined in the rest frame of the vector meson: $\theta_{\Kst}$ is the angle between the charged kaon and the $B$ meson in the \Kstar rest frame; $\theta_{\rho}$ is the angle between the positively charged (or only charged) pion and the $B$ meson in the $\rho$ rest frame.  In the analysis of the \KstOne channels, we make use of the helicity observables, defined for $\alpha = \rho,\ \Kst$ as $\calH_{\alpha} = \cos(\theta_{\alpha})$.  Occasionally, we refer to a specific charge state, e.g. \rhoz, which we indicate with the notation $\calH_{\rhoz} = \cos(\theta_{\rhoz})$.

The longitudinal polarization fraction \fL for $\Bz\ra\rho\KstOne$ can be extracted from the differential decay rate, parameterized as a function of $\theta_{\Kst}$ and $\theta_{\rho}$:
\beqa \label{eq:hel}
{1 \over \Gamma}&&
{ {d^2 \Gamma}\over {d\cos\theta_{K^{*}} d\cos\theta_{\rho}} } \propto   \\
&&{1\over 4}
(1-f_L)\sin^2\theta_{K^{*}}\sin^2\theta_{\rho}+f_L \cos^2\theta_{K^{*}} \cos^2\theta_{\rho} ~. \nonumber
\eeqa

The \CP-violating asymmetry is defined as 
\begin{equation}
\acp \equiv \frac{\Gamma^- - \Gamma^+}{\Gamma^- + \Gamma^+}  ~,
\end{equation} 
where the superscript on the decay width $\Gamma$ refers to the charge of the kaon from the \Kst decay.

All results in this paper are based on extended maximum likelihood (ML) fits as described in Section \ref{sec:mlfit}.  In each analysis, loose criteria are used to select events likely to contain the desired signal $B$ decay (Sec.~\ref{sec:eventsel}-\ref{sec:finalsel}).  A fit to kinematic and topological discriminating variables is used to differentiate between signal and background events and to determine signal event yields, $CP$-violating asymmetries, and longitudinal polarization fractions, where appropriate.  In all of the decays analyzed, the background is dominated by random particle combinations in continuum ($e^+e^-\to q{\bar q}$, $q=u,d,s,c$ ) events.  Although \qq\ background dominates the selected data sample, background from other \BB\ decays tends to have more signal-like distributions in the discriminating variables.  The dominant \BB\ backgrounds are accounted for separately in the ML fit, as discussed in Sec.~\ref{sec:BBbkg}.  Signal event yields are converted into branching fractions via selection efficiencies determined from Monte Carlo (MC) simulations of the signal as well as auxiliary studies of the data.

%-------------------------------------------------
\section{Detector and Data} \label{sec:detector}
%-------------------------------------------------

For this analysis we use the full \babar\ dataset, collected at the PEP-II asymmetric-energy \epem\ collider located at the SLAC National Accelerator Laboratory.  The dataset consists of \nbb \BB\ pairs originating from the decay of the \FourS\ resonance, produced at a center-of-mass (CM) energy $\sqrt{s} = 10.58$ GeV.  This effectively doubles the dataset from the previous \babar\ measurement~\cite{babar_rhoKst2006}.  

The asymmetric beam configuration in the laboratory frame provides a boost to the $\Upsilon(4S)$ of $\beta\gamma = 0.56$.  This results in a charged particle laboratory momentum spectrum from $B$ decays with an endpoint near 4 \gev. Charged particles are detected and their momenta measured by the combination of a silicon vertex tracker, consisting of five layers of double-sided detectors, and a 40-layer central drift chamber, both operating in the 1.5-T magnetic field of a solenoid.  For charged particles within the detector acceptance, the average detection efficiency is in excess of 96\% per particle.  

Photons are detected and their energies measured by a CsI(Tl) electromagnetic calorimeter (EMC).  The measured $\pi^0$ mass resolution for $\piz$'s with laboratory momentum in excess of 1 \gev\ is approximately 8 MeV.

Charged particle identification (PID) is provided by the average energy loss in the tracking devices and by an internally reflecting ring-imaging Cherenkov detector (DIRC) covering the central region.  Additional information that we use to identify and reject electrons and muons is provided by the EMC and the detectors installed in a segmented solenoid flux return (IFR).
The \babar\ detector is described in detail in Ref.~\cite{BABARNIM}.

%-----------------------------------------------------------
\section{Candidate Reconstruction and \boldmath{B} Meson Selection}
\label{sec:eventsel}
%-----------------------------------------------------------

We reconstruct $B$-daughter candidates through their decays $\rhoz\ra\pip\pim$, $\fz\ra\pip\pim$, $\rhom\ra\pim\piz$, $\Kstz\ra\Kp\pim$, $\Kstp\ra\Kp\piz$, and $\piz\ra\gaga$.  We apply the same selection criteria for \fz\ and \rhoz\ candidates.  

The \KstOne channels are analyzed separately from the \KstZero and \KstTwo decays, though the analyses share many similarities, including most event selection requirements.  Where the analyses differ, we specify the \KstOne channels as the ``low mass region'' (LMR), distinguished by the $K\pi$ mass requirement of $750 < \mkpi < 1000$ \mev.  The \KstZero and \KstTwo analyses are performed in the ``high mass region'' (HMR), $1000 < \mkpi < 1550$ \mev.

The invariant masses of the $B$-daughter candidates must satisfy the following requirements: $120 < m_{\gaga} < 150$ MeV, and either $470 < \mpipi < 1070$ MeV (LMR) or $470 < \mpipi < 1200$ MeV (HMR).  The $\pi\pi$ and $K\pi$ mass intervals are chosen to include sidebands large enough to parameterize the backgrounds.  

All photons are required to appear as a single cluster of energy in the EMC, not matched with any track, and to have a maximum lateral moment of 0.8.  We require the energy of the photons to be greater than 50 MeV and the \piz energy to be greater than 250 MeV, both in the laboratory frame.  
All charged tracks are required to originate from within $10\cm$ of the beamspot in the direction along the beam axis and within $1.5\cm$ in the plane perpendicular to that axis.  Charged kaon candidates are additionally required to have at least 12 hits in the drift chamber and a transverse momentum of $p_T> 100\mev$.  The charged tracks are identified as either pions or kaons by measuring the energy loss in the tracking devices, the number of photons recorded by the DIRC, and the corresponding Cherenkov angle; these measurements are combined with information from the EMC and the IFR, where appropriate, to reject electrons, muons, and protons.  

When reconstructing \rhom\ and \Kstp\ candidates, the mass of the \piz\ candidate is constrained to its nominal value~\cite{PDG}.  The \piz\ is constrained to originate from the interaction point, taking into account the finite $B$ meson flight distance; the charged track is required to originate from the interaction point.  For \rhoz\ and \Kstz\ candidates, the two charged tracks are required to originate from a common vertex, as determined by a generalized least squares minimization using Lagrange multipliers; we require the change in $\chi^2$ between two successive iterations in the fitter to be less than 0.005, with a maximum of 6 iterations.  The $B$ meson candidate is formed by performing a global Kalman fit to the entire decay chain.

A $B$-meson candidate is characterized kinematically by the energy-substituted mass \mes\ and the energy difference \DE, defined in the \UfourS\ frame as 
\begin{eqnarray}
\mes &=& \sqrt{{1\over4}s - \pvec_B^{*2}} \qquad \mbox{and}  \nonumber \\
\DE  &=& E_B^*-\half\sqrt{s} \ , \nonumber 
\end{eqnarray}
where $q_B^*=(E_B^*,\pvec_B^*)$ is the four momentum of the $B$-candidate in the \UfourS\ frame and $s$ is the square of the invariant mass of the electron-positron system.  \mes\ and \DE\ are favorable observables because they are nearly uncorrelated.  The small correlation is accounted for in the correction of the fit bias (see Sec.~\ref{sec:validation}).  
Correctly reconstructed signal events peak at zero in \DE\ and at the \B\ mass~\cite{PDG} in \mes, with a resolution in \mes\ of around 2.5 MeV and in \DE\ of 17-37 MeV.  We select events with $5.26 < \mes < 5.2893$ GeV.  For \frhozKstz, we require $|\DE| < 0.10$ GeV, while for \frhomKstp, we allow $-0.17 < \DE < 0.10\gev$ to account for a long low-side tail resulting from poorly reconstructed \piz's.

%--------------------------------------------------------
\section{Sources of Background and Suppression Techniques} 
\label{sec:bkg}
%--------------------------------------------------------

Production of $B{\bar B}$ pairs accounts for only about 25\% of the total hadronic cross section at the \UfourS\ peak.  The bulk of the cross section arises from continuum events.  Tau-pair production and other QED processes contribute as well.  We describe below the main sources of background and discuss techniques for distinguishing them from signal.

\subsection{QED and tau-pair backgrounds}

Two-photon processes, Bhabha scattering, muon- and tau-pair production are characterized by low charged track multiplicities.  Bhabha and muon-pair events are significantly prescaled at the trigger level.  We further suppress these and other tau and QED processes via a minimum requirement on the event track multiplicity.  We require the event to contain at least one track more than the topology of our final state. These selection criteria are more than 90\%  efficient when applied to signal.  From Monte Carlo simulations~\cite{geant} we determine that the remaining background from these sources is negligible.

\subsection{QCD continuum backgrounds}

The dominant background arises from random combinations of particles in continuum $\epem\ra\qqbar$ events ($q=u,d,s,c$). The angle $\theta_T$ between the thrust axis \cite{thrust} of the $B$ candidate in the \FourS\ rest frame and that of the remaining particles in the event is used to suppress this background.  Jet-like continuum events peak at values of $|\costhr|$ close to 1, while spherical \B\ decays exhibit a flat distribution for this variable.  We require that events satisfy $|\costhr|<0.7$. 

Further rejection is achieved by restricting the range of the helicity angle $\calH$ of the $\rho$ and \Kst\ mesons (see Fig.~\ref{fig:helicity_angles}).  We require $|\calH_{\rhoz}|<0.9$, $-0.8 < \calH_{\rhop} < 0.9$, $-0.85 < \calH_{\Kstz} < 1.0$, and $-0.8 < \calH_{\Kstp} < 1.0$.  These requirements reject regions of phase space with low momentum \pip's and \piz's, where backgrounds are typically large.

Additional separation of signal and background is provided by a Fisher discriminant \xf\ exploiting four variables sensitive to the production dynamics and event shape: the polar angles (with respect to the beam axis in the \epem\ CM frame) of the $B$ candidate momentum and of the $B$ thrust axis; and the zeroth and second angular moments $L_{0,2}$ of the energy flow, excluding the $B$ candidate. The moments are defined in the CM frame by 
\begin{equation}
L_j = \sum_i p_i \left|\cos\theta_i\right|^j ~,
\end{equation}
where $i$ labels a track or EMC cluster, $\theta_i$ is its angle with respect to the $B$ thrust axis, and $p_i$ is its momentum.  

We find that \xf\ in continuum background is mildly correlated with the tagging category~\cite{Btagging}, which identifies the flavor of the other $B$ in the event and places it into one of six categories based upon how it is identified.  Although the tagging category is not used elsewhere in this analysis, we find that the overall signal-to-background separation provided by \xf\ can be slightly improved by removing this correlation.  For each tagging category as well as the category for which no $B$ tag is assigned, we fit the \xf\ distribution with a Gaussian with different widths above and below the mean.  We then shift the mean of the \xf\ distribution in each tagging category to align it with the average value of the \xf\ means in all tagging categories.  
The \xf\ distributions typically have a mean around $-0.25$ with an average width around $0.45$; shifts are less than $0.03$ for all categories except for the lepton-tagged events (the tagging category with the highest purity), for which the shift is about $0.35$.  The Fisher variable provides about one standard deviation of discrimination between $B$ decay events and continuum background.

\subsection{$B\ra$ charm backgrounds}

We suppress the background from $B$ mesons decaying to charm by forming the invariant mass $m_{D}$ from combinations of two or three out of the four daughter particles' four-momenta.  For \frhozKstz, we consider $D$ candidates decaying to $\Km\pip$ and $\Km\pip\pip$.  For \frhomKstp, we consider the combinations $\Km\pip$ and $\Km\pip\piz$.  The event is retained only if $|m_{D}-m_{D}^{PDG}| > 40$ MeV for all cases except for the $D$ meson formed with $\Kp\pim$ in the \frhomKstp channel, where we require $|m_{D}-m_{D}^{PDG}| > 20$ MeV; $m_{D}^{PDG}$ is the nominal \Dp or \Dz meson mass~\cite{PDG}.

These $D$ vetoes greatly reduce the amount of $B\ra$ charm background in our samples, but as many of these channels have large branching fractions $\calO(10^{-1}-10^{-3})$, we include several charm backgrounds as separate components of the maximum likelihood fit, as detailed in Sec.~\ref{sec:BBbkg}.

\subsection{\BB\ backgrounds} \label{sec:BBbkg}

Although the dominant background arises from continuum \qqbar\ events, care must be taken to describe the backgrounds from other $B$ decays, as they have more signal-like distributions in many observables.  For \frhozKstzOne, we consider seven \BB\ background categories:  \rhozKstzZero; \fzKstzZero; \ftwoKstzOne with $\ftwo\ra\pip\pim$; \aonemKp with $\aonem\ra\rhoz\pim$; $\Bz\ra\Dm\pip$ with $\Dm\ra\Kp\pim\pim$; a combination of three $B\ra\Dzb X$ channels with $\Dzb\ra\Kp\pim\piz$; and a branching fraction-weighted combination of 13 other dominant charmless $B$ decay channels (charmless cocktail), which have a high probability of passing our selection.  The dominant channels in the charmless cocktail are $\Bp\ra a_1^0 \Kp$ with $a_1^0\ra\rhom\pip$ and $\Bp\ra\etapr\Kp$ with $\etapr\ra\rhoz\gamma$.  Most channels in the cocktail include a real \rhoz\ or \Kstz.  The number of expected events in each category is given in Table~\ref{tab:BBbkg_00}.

\begin{table}[tbh]
\caption{
\BB\ background categories for \rhozKstz and expected yields in the LMR and HMR.
}
\label{tab:BBbkg_00}
\vspace{-1em}
\begin{center}
\begin{tabular}{lcc}
\dbline
\frhozKstz background	& LMR		& HMR	\\
\sgline
\rhozKstzZero 		&$215\pm 34$	&---	\\
\fzKstzZero		& $19\pm  6$	&---	\\
\ftwoKstzOne		& $47\pm  3$	&---	\\
\aonemKp		& $15\pm  3$	&$  40 \pm ~9$	\\
$\Bz\ra\Dm\pip$		&$209\pm 10$	&$ 922 \pm 45$	\\
$B\ra\Dzb X$		&$433\pm 23$	&$1798 \pm 83$	\\
charmless cocktail	& $76\pm 22$	&$ 149 \pm 34$	\\
\dbline
\end{tabular}
\end{center}
\end{table}

For the \KstzZero and \ffzKstzTwo signals, the background categories are the same, except that the \KstzOne replaces the \KstzZero in the first two background categories.  As will be described in Sec.~\ref{sec:mlfit_hmr}, the first stage of the fit to the HMR is insensitive to the composition of the $\pip\pim$ mass spectrum; therefore \frhozKstzOne, \ffzKstzOne, and \fftwoKstzOne are included in the same \KstzOne category.  Additionally, due to the wider $\Kp\pim$ mass range in the HMR, 28 charmless $B$ decay channels are combined in the charmless cocktail.  

In analyzing \frhomKstpOne, we consider four \BB\ background categories: \rhomKstpZero, \aonemKp with $\aonem\ra\rhom\piz$, $\Bz\ra\rhop\rhom$, and $\Bm\ra\Dzb\rhom$ with $\Dzb\ra\Kp\pim\piz$.  The number of expected events in each category is given in Table~\ref{tab:BBbkg_mp}.  For the HMR, \frhomKstpOne replaces the signal mode \frhomKstpZero as a background; the other categories remain the same.  

\begin{table}[tbh]
\caption{
\BB\ background categories for \rhomKstp and expected yields in the LMR and HMR.
}
\label{tab:BBbkg_mp}
\vspace{-1em}
\begin{center}
\begin{tabular}{lcc}
\dbline
\frhomKstp background	& LMR		& HMR	\\
\sgline
\rhomKstpZero 		&$60 \pm 23$	&---	\\
\aonemKp		&$ 7 \pm 2$	&$13 \pm 3$	\\
$\Bz\ra\rhop\rhom$	&$ 9 \pm 1$	&$15 \pm 2$	\\
$\Bm\ra\Dzb\rhom$	&$129 \pm 17$	&$427 \pm 58$	\\
\dbline
\end{tabular}
\end{center}
\end{table}

In the HMR fits, the \KstOne yields are allowed to float.  The HMR \KstZero yields are extrapolated into the LMR using a ratio of LMR to HMR MC efficiencies; these \KstZero background yields are then fixed in the LMR fits.  The \ftwoKstzOne background yield is determined using a high $m_{\pip\pim}$ sideband, as described in Sec.~\ref{sec:ftwo}; this yield is fixed in the \frhozKstzOne fit.  

All other \BB\ backgrounds are modeled from the simulation, with yields fixed to experimentally measured branching fraction \calB\ values~\cite{PDG}.  For a few channels entering the charmless cocktail, no \calB\ measurements exist; in those cases, theory predictions are combined with other estimates and a 100\% uncertainty is assigned to the branching fractions.  These unmeasured charmless channels account for approximately 26\% of the charmless cocktail background in the LMR and 40\% in the HMR (see Table~\ref{tab:BBbkg_00}).  Uncertainties on the \BB\ branching fractions are accounted for as systematic uncertainties (see Sec.~\ref{sec:syst}).

%---------------------------------------------------
\section{Final Sample Criteria}\label{sec:finalsel}
%---------------------------------------------------

After all selection criteria discussed in Sec.~\ref{sec:eventsel}-\ref{sec:bkg} have been applied, the average number of combinations per event in data is 1.02 for $\rhoz/\fz \Kstz$ and 1.16 for \frhomKstp.  We select the candidate with the highest $\chi^2$ probability in a geometric fit to a common \B\ decay vertex. In this way the probability of selecting the correctly reconstructed event is a few percent higher with respect to a random selection.

 The sample sizes for the decay chains reported here range from 9700 to $37\,000$ events, where we include sidebands in all discriminating variables (except the helicities) in order to parameterize the backgrounds.

%---------------------------------------------------
\section{Maximum Likelihood Fit}\label{sec:mlfit}
%---------------------------------------------------

The candidates that satisfy the selection criteria described in Secs.~\ref{sec:eventsel}--\ref{sec:finalsel} are subjected to an unbinned, extended maximum likelihood fit to extract signal yields.  In all fits, the signal and \BB\ background components are modeled with a Monte Carlo simulation of the decay process that includes the response of the detector and reconstruction chain~\cite{geant}. 

\subsection{Low mass region fit}

In the low mass region, we obtain the yields, charge asymmetries \acp, and longitudinal polarization fractions \fL\ from extended maximum likelihood fits to the seven observables: \DE, \mes, \xf, and the masses and helicities of the two resonance candidates ($m_{\pi\pi}$, $m_{K\pi}$, $\calH_{\rho}$, and $\calH_{\Kst}$).  The fits distinguish among several categories: \qqbar\ background, \BB\ background (see Sec.~\ref{sec:BBbkg}), and signal.  The signals \frhozKstzOne and \ffzKstzOne are fit simultaneously.  For each event $i$ and category $j$ we define the probability density functions (PDFs) $\calP^i_j$ as
\begin{eqnarray}
\calP^i_j  &=& \calP_j(\mes^i) \calP_j(\DE^i) \calP_j(\xf^i) \label{eq:pdf} \\ 
&&\calP_j(m^i_{\pi\pi}) \calP_j(m^i_{K\pi}) \calP_j({\calH^i_{\rho}}) \calP_j({\calH^i_{\Kst}})\, , \nonumber
\end{eqnarray}
with the resulting likelihood $\cal L$:
\begin{eqnarray}
{\cal L} &=& \frac{e^{-\sum_j Y_j}}{N!} \prod_{i=1}^N \sum_j Y_j \calP_j^i\ , \label{eq:totalL}
\end{eqnarray}
where $Y_j$ is the fitted yield for category $j$ and $N$ is the number of events entering the fit.  For the \rhoz/\fz analysis, we use the absolute value of $\calH_{\rho}$ in the fit, as the distribution is symmetric.  We split the yields by the flavor of the decaying $B$ meson in order to measure \acp.  We find correlations among the observables to be occasionally as high as $30$\% in simulations of the \BB\ backgrounds, whereas they are small in the data samples, which are dominated by \qqbar\ background.  In signal, correlations are typically less than 1\% and occasionally as large as 14\%.  Correlations amongst observables are accounted for by evaluating the fit bias (see Sec.~\ref{sec:validation}).

\subsection{High mass region fit} 
\label{sec:mlfit_hmr}

In the high mass region, the ML fit uses the five observables: \DE, \mes, \xf, $m_{\pi\pi}$, and $m_{K\pi}$.  For \frhomKstpZero, these five observables are combined in an extended ML fit, as above.  

For the $\rhoz/\fz\KstzZero$ and \ffzKstzTwo channels, we perform the ML fit in two stages.  Due to the potential complexity of the resonant and non-resonant structures in the $\pip\pim$ and $\Kp\pim$ invariant mass spectra, as well as the fact that many of these structures are quite broad, non-trivial correlations exist between several of the ML fit hypotheses.  Attempts to perform the fit in a single stage using simulated data (see Sec.~\ref{sec:validation} for the general procedure) demonstrate unacceptable convergence rates in some scenarios.  Removing $m_{\pi\pi}$ from the ML fit greatly improves the convergence rates.  We therefore employ a two-stage procedure for these HMR fits.  In the first step, we perform an ML fit using only \DE, \mes, \xf, and $m_{K\pi}$; this allows us to separate out ``inclusive" \KstzZero\ and \KstzTwo\ signal from \qqbar\ and \BB\ backgrounds.  If we observe sufficient (greater than $3\sigma$ statistical significance) signal in the ``inclusive" \Kstz\ channels, we perform a second-stage ML fit to $m_{\pi\pi}$ for selected signal events.  Technical details are given below.

The PDF for the first-stage fit can be written as
\begin{equation}
\calP^i_j  = \calP_j(\mes^i) \calP_j(\DE^i) \calP_j(\xf^i) \calP_j(m^i_{K\pi}) \label{eq:pdf_HMR}
\end{equation}
for event $i$ and category $j$.  

In the event of significant signal in the ``inclusive'' \KstzZero\ or \KstzTwo\ channels, we apply the {\em sPlot} technique~\cite{sPlots} to the results of this first fit, which allows us to calculate a weight value for each event in each category (signal, \BB\ background, etc.) based upon the covariance matrix from the likelihood fit and the value of the PDF for that event.  Specifically, the {\em sWeight} for event $i$ of category $n$ is given by
\begin{equation}
w_n^i = \frac{\sum_{j=1}^{N_c} V_{nj} \calP^i_j}
	{\sum_{k=1}^{N_c} Y_k \calP^i_k} ~,
\end{equation}
where $N_c$ is the number of categories in the fit, $V_{nj}$ is the covariance matrix element for categories $n$ and $j$, and $Y_k$ is the yield of category $k$.

The {\em sWeight} for a given event indicates how much that event contributes to the total yield in that category; {\em sWeights} can be less than zero or greater than one, but the sum of all {\em sWeights} for a given category reproduces the ML fit yield for that category.  

The {\em sWeights} from this procedure are used to create two datasets: the {\em sWeighted} \KstzZero and \KstzTwo signal samples.  These weighted datasets allow us to determine the $\pip\pim$ mass distribution for the two signal samples of interest; these {\em sPlots} are faithful representations of $m_{\pipi}$ for the \KstzZero and \KstzTwo signal components, assuming no correlation between $m_{\pipi}$ and the observables used to generate the {\em sWeights}.  For signal MC, we find a maximum correlation of 8\% between $m_{\pipi}$ and the other observables, with correlations typically less than 2\%.

In the second stage, we fit the {\em sWeighted} \KstzZero and \KstzTwo $m_{\pipi}$ distributions to \rhoz\ and \fz hypotheses.  A non-resonant $\pip\pim$ component is found to be consistent with zero.  A \sigfz component is considered in studies of systematic uncertainties (see Sec.~\ref{sec:syst}).  This fit gives us the final signal yield for the \frhozKstzZero, \ffzKstzZero, and \ffzKstzTwo channels.  This procedure also determines the \frhozKstzTwo yield but, as we do not include helicity information in the fit, we cannot measure \fL, and thus we consider that channel a background.

Due to the two-stage nature of the $\rhoz/\fz\KstzZero$ and \ffzKstzTwo fits, the statistical uncertainty has two components.  The first is from the uncertainty on the $m_{\pi\pi}$ fit to extract the fraction of \rhoz/\fz events in the {\em sWeighted} sample.  The second is a fraction of the uncertainty on the ``inclusive'' \KstzZero (\KstzTwo) yield, the coefficient of which is given by the ratio of \rhoz\ or \fz events to the total number of inclusive \KstzZero (\KstzTwo) signal events.

%---------------------------------------
\section{Signal and background model}
\label{sec:pdfparms}
%---------------------------------------

PDF shapes for the signals and \BB\ backgrounds are determined from fits to MC samples.  For the \qqbar\ category we use data sidebands, which we obtain by excluding the signal region.  To parameterize the \qq\ PDFs for all observables except \mes, we use the sideband defined by $\mes< 5.27$\gev; to parameterize \mes, we require $|\DE|>0.06$\gev for \frhozKstz or $\DE<-0.12$ and $\DE>0.08$\gev for \frhomKstp.  The excluded \DE\ region is larger for \frhomKstp due to the poorer \DE\ resolution resulting from having two \piz's in the final state.

Signal events selected from the MC contain both correctly and incorrectly reconstructed $B$-meson candidates; the latter are labeled ``self-crossfeed'' (SXF). SXF occurs either when some particles from the correct parent $B$ meson are incorrectly assigned to intermediate resonances or when particles from the rest of the event are used in the signal $B$ reconstruction.  The fraction of SXF events ranges from 2--7\% for $\rhoz/\fz\Kstz$ candidates and from 13--22\% for \frhomKstp candidates.  We include both correctly reconstructed and SXF signal MC events in the samples used to parameterize the signal PDFs.

We use a combination of Gaussian, exponential, and polynomial functions to parameterize most of the PDFs. For the \mes\ distribution of the \qqbar\ background component, we use a parameterization motivated by phase-space arguments \cite{argus}.  

In the \KstOne (LMR) fits, the following observables are free to vary: the signal yields, longitudinal fraction \fL\ for $\rho\KstOne$, and signal charge asymmetries \acp; the \qqbar\ background yields and background \acp; and the parameters that most strongly influence the shape of the continuum background (the exponent of the phase-space-motivated \mes\ function; dominant polynomial coefficients for \DE, resonance masses, and helicities; fraction of real $\rho$, \fz, and \Kst\ resonances in the background; and the mean, width, and asymmetry of the main Gaussian describing \xf).  For the HMR fits, the equivalent parameters are allowed to float, except no \fL or \acp\ parameters are included, and the \KstOne background yields are floated.

\subsection{LASS parameterization of \KstZero}
\label{sec:LASS}

The $J^P = 0^+$ component of the $K\pi$ spectrum, which we denote \KstZero, is poorly understood; we generate MC using the LASS parameterization~\cite{LASS,Latham}, which consists of the \KstZeroRes\ resonance together with an effective-range non-resonant component.  The amplitude is given by
\begin{eqnarray}
{\mathcal A}(m_{K\pi}) &=& \frac{m_{K\pi}}{q \cot{\delta_B} - iq} \\ &+& e^{2i \delta_B} 
\frac{m_0 \Gamma_0 \frac{m_0}{q_0}}
     {(m_0^2 - m_{K\pi}^2) - i m_0 \Gamma_0 \frac{q}{m_{K\pi}} \frac{m_0}{q_0}} ~,
\nonumber \\
\cot{\delta_B} &=& \frac{1}{aq} + \half r q ~,
\end{eqnarray}
where $m_{K\pi}$ is the $K\pi$ invariant mass, $q$ is the momentum of the $K\pi$ system, and $q_0 = q(m_0)$.  We use the following values for the scattering length and effective-range parameters: $a=2.07\pm0.10\,(\gev)^{-1}$ and $r=3.32\pm0.34\,(\gev)^{-1}$~\cite{Latham}.  For the resonance mass and width we use $m_0=1.412\gev$ and $\Gamma_0=0.294\gev$.  

In the HMR, we parameterize the $m_{K\pi}$ distribution of the \KstZero signal category with a Gaussian convolved with an exponential.  This shape reasonably approximates the LASS distribution, given the limited statistics in this analysis, and is chosen to reduce computation time.  In the LMR, we use a linear polynomial, as only the tail of the \KstZero enters the LMR $m_{K\pi}$ region.

\subsection{PDF corrections from data calibration samples}
\label{sec:control}

The decays $\Bz\ra\Dm\pip$ ($\Dm\ra\Kp\pim\pim$) and $\Bz\ra\Dzb\piz$ ($\Dzb\ra\Kp\pim\piz$) have the same particle content in the final state as the signal, as well as large branching fractions. They are used as calibration channels.  We apply the same selection criteria described in Secs.~\ref{sec:eventsel}--\ref{sec:finalsel}, except that the $m_{\pi\pi}$ and $m_{K\pi}$ mass restrictions are replaced with $1.85 < m_{\Dm} < 1.89$ GeV or $1.83 < m_{\Dzb} < 1.89$ GeV and no $D$ meson veto is applied. We use the selected data to verify that the ML fit performs correctly and that the MC properly simulates the \xf, \DE, and \mes\ distributions.  From these studies, we extract small corrections to the MC distributions of \DE\ and \mes, which we apply to the signal PDFs in our LMR and HMR likelihood fits.  We find that it is not necessary to correct the PDF for \xf.

%----------------------------------------------------------
\section{Background \ftwoKstzOne yield from high $m_{\pip\pim}$ sideband}
\label{sec:ftwo}
%----------------------------------------------------------

\begin{figure}[!htbp]
\begin{center}
  \includegraphics[width=1.0\linewidth]{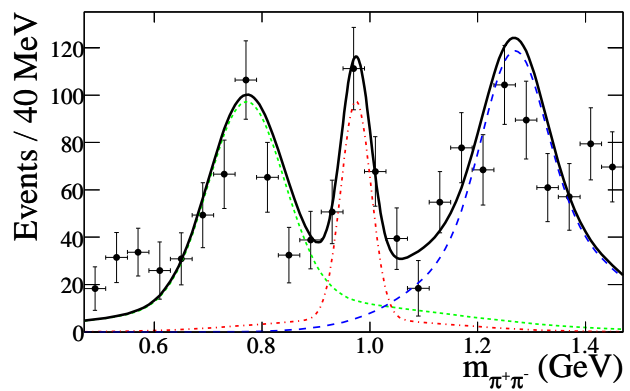}
  \caption{(color online)
    $\pip\pim$ mass spectrum for {\em sWeighted} \KstzOne events.  
    The solid curve is the fit function, the [green] dotted curve is \rhoz, [red] dash-dotted is \fz, and [blue] dashed is \ftwo.
   }
  \label{fig:proj_HighMass_pipi}
\end{center}
\end{figure}

To extract the \fftwoKstzOne yield, we select the LMR for $m_{K\pi}$ and require a $\pip\pim$ invariant mass within the range $0.47<\mpipiC<1.47$\gev.  We perform an ML fit with the observables \DE, \mes, \xf, and $m_{K\pi}$, and create a dataset of {\em sWeighted} \KstzOne events.  We then fit the \mpipiC spectrum of the {\em sWeighted} \KstzOne events to \rhoz, \fz, and \ftwo hypotheses (see Fig.~\ref{fig:proj_HighMass_pipi}).  We find \nftwoKstz \fftwoKstzOne events after subtracting a $25\pm 13$ event fit bias, which includes systematics; see Sec.~\ref{sec:validation} for details of the fit bias estimation method.  The MC efficiency $\epsilon$ of \ftwoKstzOne is 11.8\% (longitudinal polarization) and 20.4\% (transverse polarization).

Note that the three-component fit in Fig.~\ref{fig:proj_HighMass_pipi} well describes the resonances of interest, though the fit quality is poor at the lowest and highest $\pip\pim$ masses.  As this study is intended to estimate the effect of higher resonances feeding into the nominal fit region, we determine that the three-component fit is sufficient.  The excess of events in the low mass region could suggest the presence of a $\sigma/f_0(600)$ resonance; this is accounted for in a separate systematic study for the nominal fit.  The excess in the highest bins could be explained by contributions from additional higher-mass resonances.  As such resonances are unlikely to affect the \rhoz\ and \fz\ yields, we leave further understanding of these resonances for future studies.

Using the MC efficiency for \fftwoKstzOne in the LMR region, which includes a tighter cut on $m_{\pi\pi}$, and assuming $\fL=0.5$, we determine that there are $47\pm 3$ \fftwoKstzOne events expected in the LMR $\rhoz/\fz\KstzOne$ fit, as indicated in Table~\ref{tab:BBbkg_00}.

\section{Fit Validation}
\label{sec:validation}

Before applying the fitting procedure to the data, we subject it to several tests.  Internal consistency is verified by performing fits to ensembles of simulated experiments.  From these we establish the number of parameters associated with the \qq\ PDF shapes that can be left free to float.  Ensemble distributions of the fitted parameters verify that the generated values are reproduced with the expected resolution.

We investigate possible biases on the fitted signal yield $Y_0$, as well as on \fL for the $\rho\KstOne$ channels, due to neglecting correlations among discriminating variables in the PDFs, as well as from cross-feed from the \BB\ background modes.  To determine these biases, we fit ensembles of experiments into which we embedd the expected number of signal and \BB\ background events randomly extracted from detailed MC samples in which correlations are fully modeled.  As correlations among fit variables are negligible for \qqbar\ events, these events are generated from the PDFs.  Each such experiment has the same number of signal and background candidates as the data.  The measured biases are given in Table~\ref{tab:results}.  In calculating the branching fractions, we subtract the bias and include a systematic uncertainty (see Sec.~\ref{sec:syst}) associated with the procedure.

The two-stage fit employed to determine the $\rhoz/\fz\KstzZero$ and \ffzKstzTwo yields (see Sec.~\ref{sec:mlfit_hmr}) complicates the validation procedure.  We perform the first stage of the fit (which extracts the ``inclusive'' \KstzZero and \KstzTwo yields) on ensembles of experiments, as described above.  The bias obtained from this study is split between the \rhoz\ and \fz channels based on the relative fraction of \rhoz\ or \fz events to the total number of signal events in that sample.

\section{Fit Results}
\label{sec:results}

The branching fraction for each decay chain is obtained from
\beq
\calB = \frac{Y-Y_0}{\epsilon~N_B~\prod{\calB_i}}\,,
\eeq
where $Y$ is the yield of signal events from the fit, $Y_0$ is the fit bias discussed in Sec.~\ref{sec:validation}, $\epsilon$ is the MC efficiency, $\calB_i$ is the branching fraction for the $i^{\mbox{th}}$ unstable $B$ daughter ($\calB_i$ having been set to unity in the MC simulation), and $N_B$ is the number of produced \Bz\ mesons.  The values of $\calB_i$ are taken from Particle Data Group world averages~\cite{PDG}. 
We assume the branching fractions of \FourS\ to \BpBm\ and \BzBzb\ to be the same and to each equal 50\%. As the branching fractions $\calB(\fz\ra\pi\pi)$ and $\calB(\KstZero\ra K\pi)$ are poorly known, we measure the products 
\beqa
\calB(\Bz\ra\fz\Kst)\times\calB(\fz\ra\pi\pi) ~~~\mbox{and} \nonumber \\
\calB(\Bz\ra X\KstZero)\times\calB(\KstZero\ra K\pi) ~.  \nonumber
\eeqa
We include the isospin ratios
\beqa
\frac{\Gamma(\KstzZero\ra\Kp\pim)}{\Gamma(\KstzZero\ra K\pi)} &=& \frac{2}{3} ~, \nonumber \\
\frac{\Gamma(\KstpZero\ra\Kp\piz)}{\Gamma(\KstpZero\ra K\pi)} &=& \frac{1}{3} ~, \nonumber \\
\frac{\Gamma(\fz(980)\ra\pip\pim)}{\Gamma(\fz(980)\ra\pi\pi)} &=& \frac{2}{3}  \nonumber
\eeqa
in our calculations of $\prod{\calB_i}$.  The efficiency $\epsilon$ is evaluated from the simulation.  For the \frhomKstp channels, we apply an efficiency correction to the MC of roughly 97\%/\piz.  The specific values are determined by calculating a correction as a function of the \piz lab momentum from a detailed MC simulation of the signal channel.  The correction is determined from a study of tau decays to modes with \piz's as well as a study of $\epem\ra\gamma\omega$ with $\omega\ra\pip\pim\piz$.  The results for all signal channels are collected in Table~\ref{tab:results}.

\begin{table*}[!bth]
\caption{
Signal yield $Y$ and its statistical uncertainty (see Sec.~\ref{sec:mlfit_hmr} for an explanation of the two errors on the \KstzZero and \KstzTwo yields); fit bias $Y_0$; detection efficiency $\epsilon$ for longitudinal (ln) and transverse (tr) polarizations, if appropriate; daughter branching fraction product $\prod\calB_i$; significance $S$ including systematic uncertainties; measured branching fraction \calB\ with statistical and systematic errors; 90\% C.L. upper limit (U.L.); longitudinal fraction \fL; and charge asymmetry \acp. 
In the case of \ffzKstz, the quoted branching fraction is the product of \BfzKstz.  For the $\Bz\ra X\KstZero$ channels, the quoted branching fraction is the product of $\calB(\Bz\ra X\KstZero)\times\calB(\KstZero\ra K\pi)$.  We include the isospin ratios $(\KstzZero\ra\Kp\pim) = 2/3$, $(\KstpZero\ra\Kp\piz) = 1/3$, and $(\fz\ra\pip\pim) = 2/3$.
}
\label{tab:results}
\begin{tabular}{lccccrccccc}
\dbline
Mode       & $Y$      &$Y_0$       &$\epsilon$(ln) &$\epsilon$(tr) &$\prod\calB_i$ & $S~$       &  \calB        & U.L. & \fL &  \acp \\
           & (events) & (events)   &(\%)$~$         &(\%)$~$         & (\%)$~$       &$(\sigma)$  & $(10^{-6})$ & $(10^{-6})$ &      &    \\
\dbline
\frhozKstzOne  &\nrhozKstz	&\biasrhozKstz	  & 14.3 & 25.1             & 66.7 &\srhozKstz    &\rrhozKstz  &--- &\fLrhozKstz &\ArhozKstz	\\
\frhozKstzZero &\nrhozKstzZero	&\biasrhozKstzZero&\multicolumn{2}{c}{9.6}  & 66.7 &\srhozKstzZero&\rrhozKstzZero  &---&---     &--- \\
\sgline
\ffzKstzOne    &\nfzKstz	&\biasfzKstz	  &\multicolumn{2}{c}{18.3} & 44.4 &\sfzKstz      &\rfzKstz       &---& ---     &\AfzKstz	\\
\ffzKstzZero   &\nfzKstzZero	&\biasfzKstzZero  &\multicolumn{2}{c}{12.5} & 44.4 &\sfzKstzZero  &\rfzKstzZero   &---& ---     &---	\\
\ffzKstzTwo    &\nfzKstzTwo	&\biasfzKstzTwo	  &\multicolumn{2}{c}{15.3} & 21.7 &\sfzKstzTwo   &\rfzKstzTwo    &---& ---     &---	\\
\sgline
\frhomKstpOne  &\nrhomKstp	&\biasrhomKstp	  & 4.9 & 11.2              & 33.3 &\srhomKstp    &\rrhomKstp &--- &\fLrhomKstp &\ArhomKstp	\\
\frhomKstpZero &\nrhomKstpZero	&\biasrhomKstpZero&\multicolumn{2}{c}{4.5}  & 33.3 &\srhomKstpZero&\rrhomKstpZero &$<\ulrhomKstpZero$ &---     &---	\\
\dbline 
\end{tabular}
\vspace{-5mm}
\end{table*}

For all signals obtained from a one-stage ML fit, we determine the significance of observation $S$ by taking the difference between the value of $-2\ln{\cal L}$ for the zero signal hypothesis and the value at its minimum.  For the \frhozKstzZero, \ffzKstzZero, and \ffzKstzTwo channels, the fit method does not readily provide a $-2\ln{\cal L}$ distribution, so we determine the significance assuming Gaussian uncertainties, which provides a conservative lower limit on $S$.

For the \frhomKstpZero channel, which has a significance less than $3\sigma$ including systematics, we quote a 90\% C.L. upper limit, given by the solution $\calB_{90}$ to the equation
\begin{equation}
\frac{\int_0^{\calB_{90}}{\calL}(b)db}{\int_0^\infty{\cal L}(b)db}=0.9\,,
\end{equation}
where ${\cal L}(b)$ is the value of the likelihood for branching fraction $b$.  Systematic uncertainties are taken into account by convolving the likelihood with a Gaussian function representing the systematic uncertainties.

% mES projections
\begin{figure}[!tbp]
\begin{center}
  \includegraphics[width=1.0\linewidth]{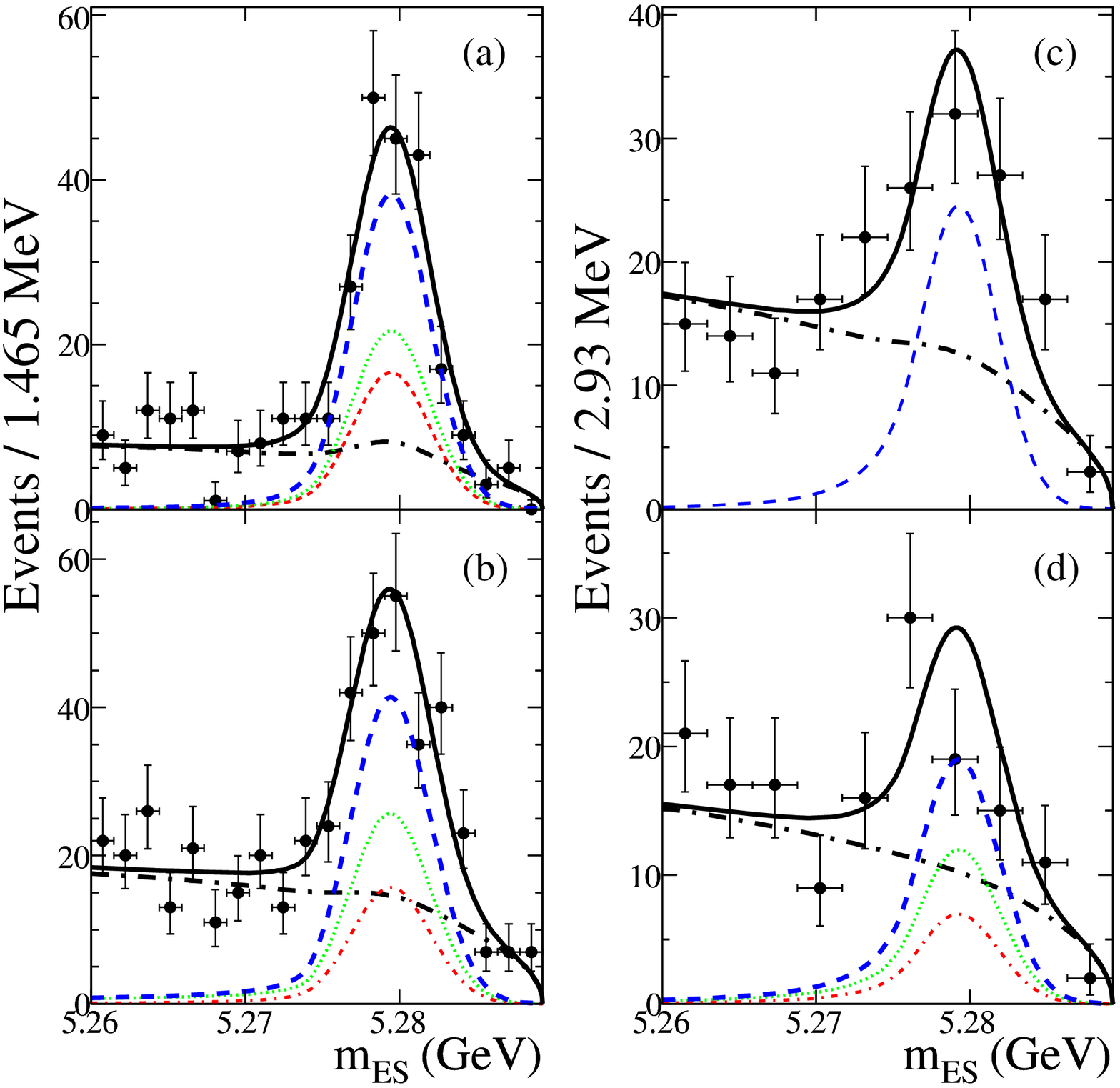}
  \caption{(color online)
$B$-candidate \mes\ projections for (a) $\rhoz/\fz\KstzOne$ (b) ``inclusive'' \KstzZero and \KstzTwo, (c) \frhomKstpOne, (d) \frhomKstpZero and \frhomKstpTwo.
    The solid curve is the fit function, [black] long-dash-dotted is the total background, and the [blue] dashed curve is the total signal contribution.  In (a) we separate the [red] dashed \rhoz\ component from the [green] dotted \fz.  In (b) and (d) \KstZero signal is [green] dotted and \KstTwo is [red] dashed.  In (b), the two-stage nature of the fit means that the \KstzZero and \KstzTwo signals include both \fz\ and \rhoz\ components, as the first stage of the HMR fit does not include information about the $\pip\pim$ mass spectrum.
   }
  \label{fig:proj_mes}
\end{center}
\end{figure}

% m(pipi) and m(Kpi) plots for K*(892)
\begin{figure}[!htbp]
\begin{center}
  \includegraphics[width=1.0\linewidth]{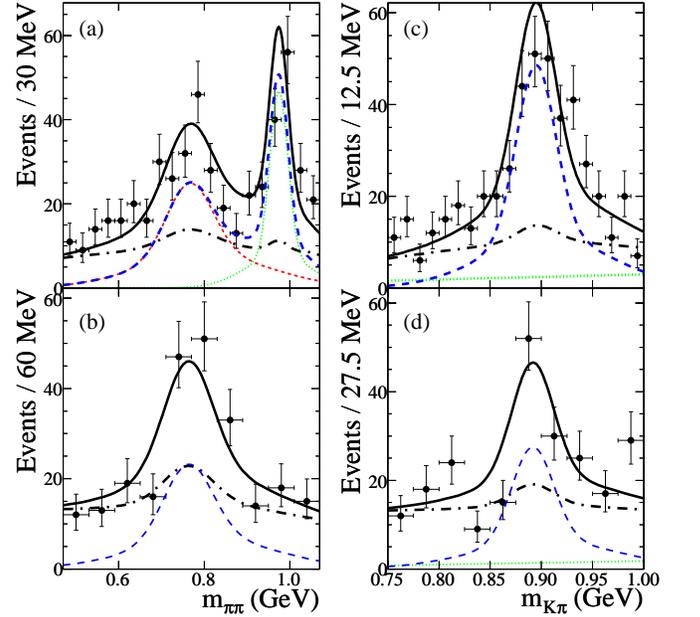}
  \caption{(color online)
Invariant mass projections for LMR (a,c) $\rhoz/\fz\KstzOne$ and (b,d) \frhomKstpOne; $\pi\pi$ mass (left) and $K\pi$ mass (right).  
    The solid curve is the fit function, [black] long-dash-dotted is the total background, and the [blue] dashed curve is the total signal contribution.  In (a) we separate the [red] dashed \rhoz\ component from the [green] dotted \fz.  In (c) and (d) \KstZero background is [green] dotted.
   }
  \label{fig:proj_mRK}
\end{center}
\end{figure}

\begin{figure}[!tbp]
\begin{center}
  \includegraphics[width=1.0\linewidth]{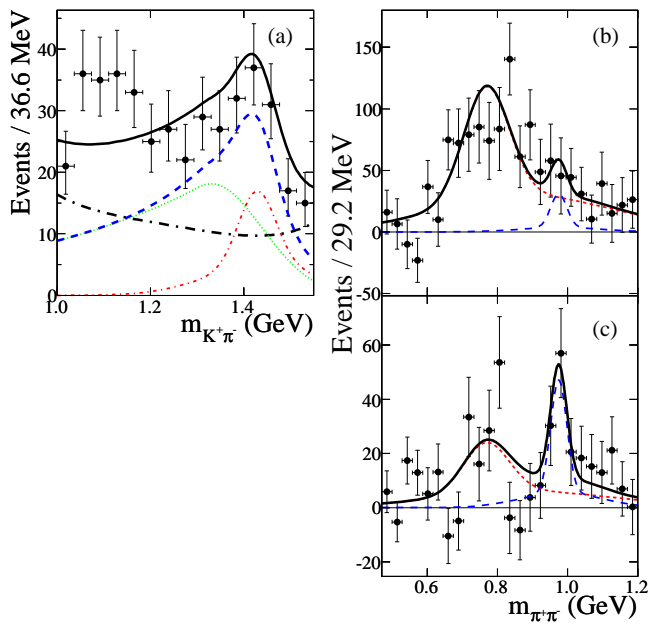}
  \caption{(color online)
Invariant mass projections for HMR \frhozKstzZero, \ffzKstzZero, and \ffzKstzTwo signals (a) $\Kp\pim$ mass, (b) $\pip\pim$ mass for {\em sWeighted} \KstzZero events, (c) $\pip\pim$ mass for {\em sWeighted} \KstzTwo events.  The solid curve is the fit function.  In (a) the [black] long-dash-dotted is the total background, the [blue] dashed curve is the total signal contribution, [green] dotted is the \KstzZero\ component, and the \KstzTwo\ is [red] dashed.  In (b) and (c) the \rhoz\ component is [red] dashed, \fz is [blue] long-dashed.
   }
  \label{fig:proj_mRK_HighMass_Kpi}
\end{center}
\end{figure}

We show in Fig.~\ref{fig:proj_mes} the data and fit functions projected onto the variable \mes, while in Fig.~\ref{fig:proj_mRK} we do the same for the $\pi\pi$ and $K\pi$ invariant masses for the LMR measurements.  
In Fig.~\ref{fig:proj_mRK_HighMass_Kpi}(a) we project the data and fit functions from the first stage of the HMR \KstzZero and \KstzTwo fits onto $m_{\Kp\pim}$.
In Figs.~\ref{fig:proj_mes}, \ref{fig:proj_mRK}, and \ref{fig:proj_mRK_HighMass_Kpi}(a) the signals are enhanced by the imposition of restrictions on the likelihood ratio, which greatly reduce the amount of background while retaining events that have a large probability to be signal.  

Figures~\ref{fig:proj_mRK_HighMass_Kpi}(b) and (c) show the results of the second-stage HMR fit, distinguishing between the \rhoz\ and \fz hypotheses.  In these plots, we do not impose a restriction on the likelihood ratio, as these {\em sWeighted} samples already contain only (b) \KstzZero or (c) \KstzTwo signal events.

Ref.~\cite{Latham} extracts the resonant \KstzZeroRes fraction of the LASS-parameterized \KstzZero\ distribution.  The resonant fraction is found to account for 81\% of the LASS shape in $\Bp\ra(K \pi)_0^{*0}\pip$ decays.  Using this resonant fraction along with the daughter branching fraction $\calB(\KstZeroRes\ra K\pi)=(93\pm 10)\%$~\cite{PDG}, we find the resonant branching fractions
\beqa
\BrhozKstzZeroRes &=& \RrhozKstzZeroRes ~;	\nonumber \\
\lefteqn{\BfzKstzZeroRes}  	\hspace{3.5cm}  \nonumber \\
                  &=& \RfzKstzZeroRes	 ~;	\nonumber \\
\BrhomKstpZeroRes &=& \RrhomKstpZeroRes	 ~,	\nonumber
\eeqa
where the uncertainties are statistical, systematic, and from the $\KstZeroRes\ra K\pi$ branching fraction, respectively.

% rho0/f0 K_0*0 and K_2*0

%--------------------------------------
\section{Systematic Uncertainties}
\label{sec:syst}
%--------------------------------------

Table~\ref{tab:systtab} summarizes our estimates of the various sources of systematic uncertainty.  We distinguish between uncertainties that concern a bias on the yield (additive) and those that affect the efficiency and total number of \BB\ events (multiplicative), since only the former affect the significance of the results.  The additive systematic uncertainties are the dominant source of systematics for the results presented in this paper.  The final row of the table provides the total systematic error in units of branching fraction for each channel.

\begin{table*}[tbh]
\caption{
Estimates of systematic uncertainties.
}
\label{tab:systtab}
\vspace{-1em}
\begin{center}
\begin{tabular}{lccccccc}
\dbline
Quantity	&\frhozKstz &\frhozKstzz &\ffzKstz &\ffzKstzz &\ffzKstzt &\frhomKstp &\frhomKstpz	\\
\sgline
Additive errors (events)&	&	&	&	&	&	&	\\
~~ML fit	   	& ~2.7	& ~3.7  & ~1.1	& ~0.3  & 0.7   & ~6.7  & 21.3  \\
~~Fit bias    		& 22.2	& 41.8  & ~1.9	& 12.5  & 5.0   & 11.9  & ~8.4  \\
~~\BB\ background    	& 14.8	& ~5.1  & ~2.7	& ~0.4  & 0.6   & ~6.0  & ~3.2  \\
~~$\fz(980)$ parameters& ~3.5	& 10.0  & ~8.8	& 10.0  & 2.5   & ---   & ---   \\
~~LASS parameters	& ---	& 29.5  & ---	& ~2.5	& 7.7	& ---   & 31.0	\\
~~Interference		& 12.2	& 57.5  & 12.9	& 6.5   & 9.9   & 8.4   & 18.1   \\
~~$\sigma/f_0(600)$	&$^{+0.0}_{-36.0}$ &$^{+0.0}_{-28.0}$ & ~1.0 & 2.1 & 6.4 & --- & --- \\
\sgline				        		
Total additive (events)	&$^{+29.7}_{-46.6}$ &$^{+77.9}_{-82.7}$  & 16.0 & 17.6	& 15.2	& 17.1	& 42.7	\\
Total additive \lbrack\bfemsix\rbrack &$^{+0.45}_{-0.71}$ &$^{+2.58}_{-2.74}$ & 0.42 & 0.67 & 1.00 & 1.24 & 6.00 \\
\sgline
Multiplicative errors (\%)&	&	&	&	&	&	&	\\
~~Track multiplicity    & 1.0	& 1.0   & 1.0	& 1.0   & 1.0   & 1.0   & 1.0   \\
~~Track finding		& 0.7	& 0.7   & 0.7	& 0.7   & 0.7   & 0.4   & 0.4   \\
~~\piz\ efficiency	& ---	& ---   & ---	& ---   & ---   & 3.8   & 3.6   \\
~~Number \BB\    	& 0.6	& 0.6   & 0.6	& 0.6   & 0.6   & 0.6   & 0.6   \\
~~Branching fractions   & --- 	& ---   & ---	& ---	& 1.2	& ---   & ---	\\
~~MC statistics    	& 0.08	& 0.07  & 0.06	& 0.04  & 0.05  & 0.06  & 0.05  \\
~~\costhr    		& 1.5	& 1.5   & 1.5	& 1.5   & 1.5   & 1.5   & 1.5   \\
~~PID			& 1.0	& 1.0   & 1.0	& 1.0   & 1.0   & 1.0   & 1.0   \\
~~\fL uncertainty	& 5.8	& ---   & ---	& ---	& ---	& 2.1   & ---	\\
\sgline 							        
Total multiplicative (\%) & 6.2	& 2.3	& 2.3	& 2.3	& 2.6	& 4.9   & 4.2	\\
\sgline
Total systematic \lbrack\bfemsix\rbrack &$^{+0.6}_{-0.8}$ &$^{+2.7}_{-2.8}$ & 0.4 & 0.7	& 1.0	& 1.3	& 6.1	\\
\dbline
\end{tabular}
\end{center}
\end{table*}

\subsection{Additive systematic errors}
\label{sec:addsyst}

\noindent{\bf ML fit}:  
We evaluate the systematic uncertainties due to the modeling of the signal PDFs by varying the relevant PDF parameters by uncertainties derived from the data control samples (see Sec.~\ref{sec:control}).  This uncertainty is larger for the \frhomKstp channels, as the $\Dzb\piz$ control sample has lower statistics than the $\Dm\pip$ sample used for the \frhozKstz channels.  

\noindent{\bf Fit bias}:
The fit bias arises mostly from correlations among the fit variables, which are neglected in the ML fit.  Studies of this bias are described in Sec.~\ref{sec:validation}.  The associated uncertainty is the sum in quadrature of half the correction and its statistical uncertainty.  For the $\rhoz/\fz\KstzZero$ and \ffzKstzTwo channels, we add the uncertainty on the total bias in quadrature with half the bias scaled by the ratio of \rhoz\ or \fz events to their sum.

\noindent{\bf \boldmath{\BB} background}: 
We estimate the uncertainty from the fixed \BB\ background component yields by repeating the fit with the yields of these components varied by their uncertainties.  For each signal channel, we add in quadrature the change in signal yield from varying each \BB\ background, and quote this as the systematic.  The uncertainty on the measured $\rho\KstZero$ branching fractions makes this a large systematic for \frhozKstzOne and \frhomKstpOne.

\noindent{\bf \boldmath{$\fz(980)$} parameters}:
The width of the \fz is not accurately measured; to account for this, we allow the mean and width of the \fz to float in the LMR fit and take half the shift in the signal yield as a systematic.  This is one of the largest systematics for \ffzKstzOne.  In the HMR, we lack the statistics to allow these parameters to float, and so perform the fit with them fixed to the parameters obtained when floating them in the LMR.  This is amongst the largest systematics for \ffzKstzZero and \frhozKstzZero.

\noindent{\bf LASS shape parameters}:
For the \KstZero channels, we vary the LASS parameters in the MC by the uncertainties listed in Sec.~\ref{sec:LASS} and re-fit the data sample with PDF parameters based on this new MC.  In each channel, we take the largest variation in the yield as a result of this procedure as the systematic.    The LASS systematic is the dominant one for \frhomKstpZero.

\noindent{\bf \boldmath{$\rhoz/\fz$} Interference}:  
The interference between the \rhoz\ and \fz integrates to zero over the symmetric $\calH_{\rho}$ range.  Additionally, the differential rate is an odd function of $\calH_{\rho}$, so the fact that we use $|\calH_{\rho}|$ in the fit means that the interference term also vanishes from the differential rate.

\noindent{\bf \boldmath{\Kst} Interference}:  
In our nominal fits, we do not account for interference between the scalar and vector \Kst, or between the vector and tensor.  We estimate the magnitude of the \Kst\ interference effect in a separate calculation, which takes into account the relevant mass and helicity acceptance functions, and varies the relative strong phases between components over the full range.  As interference can affect the \Kst\ lineshape, we conservatively take this systematic to be additive.  This is among the dominant systematic uncertainties in the HMR fits.

\noindent{\bf \boldmath{\ftwo} Interference}:  
The fit described in Sec.~\ref{sec:ftwo}, used to estimate the background from $\Bz\ra\ftwo\KstzOne$ decays, is performed without interference.  Interference terms vanish when integrating over the full solid angle, however the requirement that the helicity angle be $|\calH_{\rho}| < 0.9$ leaves a non-zero interference term.  For the case of interference between the \fz\ and \ftwo, the scalar \fz\ may interfere with the longitudinal component of the \ftwo.  Adding this term to the fit results shown in Fig.~\ref{fig:proj_HighMass_pipi}, assuming $\fL = 0.5$ for the \ftwo, and scanning over the unknown phase difference between the \fz\ and \ftwo, we find a maximum yield difference between the case of no interference of $\pm 8.4\%$ in the \fz\ region.  As the interference depends upon the sine of the unknown phase, we divide by $\sqrt{2}$ and report an additive $\fz-\ftwo$ interference systematic of 12.8 events ($5.8\%$).  Using a similar procedure for $\rhoz-\ftwo$ interference, we report a systematic of 6.8 events ($1.8\%$).  For the \ffzKstzOne\ measurement, this is the dominant systematic. 

\noindent{\bf \boldmath{\sigfz} resonance}:  The scalar $\sigma/f_0(600)$ is poorly understood and its parameters uncertain.  We estimate the effect of a possible $\sigma/f_0(600)$ resonance by including $\sigfz\Kst$ as a separate component in our fits.  We parameterize the \sigfz using a relativistic Breit Wigner function with $m=513$\mev and $\Gamma=335$\mev~\cite{sigma}.  As we lack $\sigfz\Kst$ MC, for the LMR \KstzOne fit we use the \ffzKstz PDF shapes for all variables except the $\pip\pim$ invariant mass.  We use the average $\sigfz\Kstz$ branching fraction from the three \Kstz\ channels to calculate how many $\sigfz\Kstz$ events are expected in each \Kstz\ sample; this $\sigfz\Kstz$ yield is then fixed in each fit.  We take 100\% of the resulting signal yield variation as a low-side systematic for the \rhoz\ channels (a non-zero \sigfz yield decreases the \rhoz\ yield) and conservatively consider this a two-sided systematic in the \fz channels.  This is the dominant systematic for \frhozKstzOne.

\subsection{Multiplicative systematic errors}

\noindent{\bf Track multiplicity}: 
The inefficiency of the selection requirements for the number of tracks in the event is a few percent. We estimate an uncertainty of 1\% from the uncertainty in the low-multiplicity tail of the \B\ decay model.

\noindent{\bf Track finding/efficiency}:
Studies of tau events determine that no efficiency correction is necessary for track finding and reconstruction.  The systematic uncertainty is determined by adding linearly 0.17\% per track in quadrature with an overall factor of 0.11\%.

\noindent{\bf \boldmath{\piz} reconstruction efficiency}: 
We apply an efficiency correction to the MC of roughly 97\%/\piz; the correction depends on the \piz\ momentum spectrum, so is somewhat different in different channels.  The uncertainty associated with this correction is roughly 1.5\%/\piz.  

\noindent{\bf Number of \boldmath{\BB} events}:  
A separate study~\cite{Bcounting} determines the overall uncertainty on the number of produced \BB\ pairs to be 0.6\%.

\noindent{\bf Branching fractions of decay chain daughters}: 
This is taken as the uncertainty on the daughter particle branching fractions from Ref.~\cite{PDG}.

\noindent{\bf MC statistics}: 
The uncertainty due to finite signal MC sample sizes (typically 430,000 generated events) is given in Table~\ref{tab:systtab}.

\noindent{\bf Event shape requirements}: 
Uncertainties due to the \costhr\ requirement are estimated from data control samples to be $0.05\times(1-(|\costhr|\mbox{ cut value}))$.

\noindent{\bf PID}:  
We estimate from independent samples that the average efficiency uncertainty associated with particle identification is 1.0\%.

\noindent{\bf \boldmath{\fL} uncertainty}:  
The signal yield reconstruction efficiency for $VV$ channels depends on \fL.  As a result, any systematic uncertainty on \fL translates into a systematic uncertainty on the efficiency through the following expression:
\begin{equation}
\label{eq:flsys}
\frac{\Delta\epsilon}{\epsilon} =
\frac{\epsilon_L - \epsilon_T}{f_L\epsilon_L + (1-f_L)\epsilon_T}\Delta f_L ~.
\end{equation}
The systematic error on \fL ($\Delta f_L$) is given in Table~\ref{tab:systtab_fL}.

\subsection{Charge asymmetry systematic errors}
\label{sec:chgasymsyst}

From the analysis of a variety of data control samples, the bias on \acp\ is found to be negligible for pions and --0.01 for kaons, due to differences between \Kp\ and \Km\ interactions in the detector material. We correct the fitted \acp\ by +0.01 and assign a systematic uncertainty of 0.02, mainly due to the bias correction.

\subsection{Systematic errors on \fL}
\label{sec:fLsyst}

Most systematic uncertainties cancel when calculating \fL.  We include uncertainties from the signal PDF modeling (``ML fit''), fit bias (for which we assign an uncertainty equal to 100\% of the bias added in quadrature with its uncertainty), \BB\ background yields, the \fz parameterization, and the possible existence of a \sigfz (where we take 100\% of the \fL variation when the $\sigfz\Kstz$ is fixed in the study described in Sec.~\ref{sec:addsyst}).  For \frhozKstzOne, the fit bias of $-0.045\pm 0.008$ provides a moderate uncertainty; for \frhomKstpOne, this bias is small ($-0.009\pm 0.014$).  See Table~\ref{tab:systtab_fL} for details.

\begin{table}[htb!]
\caption{ Estimates of systematic errors on $f_L$.}
\label{tab:systtab_fL}
\vspace{-1em}
\begin{center}
\begin{tabular}{lcc}
\dbline
Quantity   &\frhozKstzOne &\frhomKstpOne	\\
\sgline			
ML fit            & 0.003	& 0.012	\\
Fit bias          & 0.046	& 0.016	\\
\BB\ background   & 0.019	& 0.024	\\
$\fz(980)$ parameters& 0.004	& ---	\\
$\sigma/f_0(600)$ & 0.100	& ---	\\
\sgline			
   Total          & 0.112	& 0.031	\\
\dbline
\end{tabular}
\end{center}
\end{table}

\section{Discussion and summary of results}
\label{sec:conclusion}

We obtain the first observations of \rhozKstzZero, \ffzKstzOne, and \frhomKstpOne with greater than $5\sigma$ significance, including systematics.  We present the first evidence for \fzKstzZero and \ffzKstzTwo, which we observe with a significance of $\sfzKstzZero\sigma$ and $\sfzKstzTwo\sigma$, respectively.  All branching fraction measurements have greater than $3\sigma$ significance including systematics, except \frhomKstpZero for which we also quote a 90\% C.L. upper limit.  No significant direct \CP-violation is observed.  Our results are consistent with and supersede those reported in Ref.~\cite{babar_rhoKst2006}. 

For the $\Kst(892)$ channels, we find the following results
\beqa
\BrhozKstz	  &=& \RrhozKstz  ~;	\nonumber \\
\fL (\frhozKstz)  &=& \fLrhozKstz ~;	\nonumber \\
\acp (\frhozKstz) &=& \ArhozKstz  ~;	\nonumber \\
					\nonumber \\
\BrhomKstp 	  &=& \RrhomKstp  ~;	\nonumber \\
\fL (\frhomKstp)  &=& \fLrhomKstp ~;	\nonumber \\
\acp (\frhomKstp) &=& \ArhomKstp  ~;	\nonumber \\
					\nonumber \\
\lefteqn{\BfzKstz}  \hspace{3cm}	\nonumber \\
		  &=& \RfzKstz	 ~;	\nonumber \\
\acp (\ffzKstz)   &=& \AfzKstz	 ~.	\nonumber
\eeqa
The \frhozKstzOne results agree with previous \babar~\cite{babar_rhoKst2006} and Belle~\cite{BelleRhoKst} measurements and are consistent with predictions from QCDF~\cite{QCDF}.  The \frhomKstpOne results are consistent with the previous \babar\ upper limit and agree with QCDF predictions.  Both the \frhozKstzOne and \frhomKstpOne branching fractions are, however, higher than the values predicted by QCDF.  We find a branching fraction for \ffzKstzOne within the previous \babar\ 90\% C.L. upper limit ($6.5\times10^{-6}$~\cite{babar_rhoKst2006}) and somewhat above the Belle limit ($3.3\times10^{-6}$~\cite{BelleRhoKst}), where we have scaled the published limits by a factor of $3/2$, as the previous analyses assumed $\calB(\fz\ra\pip\pim)=100\%$ whereas this measurement includes the isospin ratio $\Gamma(\fz\ra\pip\pim)/\Gamma(\fz\ra\pi\pi) = 2/3$.  The \ffzKstzOne branching fraction result is within one sigma of the QCDF prediction of $4.8^{+5.3}_{-2.0}\times10^{-6}$, which is scaled by a factor of $3/4$, as Ref.~\cite{QCDF} assumes $\calB(\fz\ra\pip\pim)=0.5$.  We note that a previous \babar\ study of $B\ra\phi\Kst$~\cite{BaBarPhiKst} observed an excess of $\Bz\ra(\Kp\Km)_0 \KstzOne$ events, where the scalar $(\Kp\Km)_0$ could include $\fz(980)$ decays.  If we assume all the observed $(\Kp\Km)_0$ excess to be from $\fz\ra\Kp\Km$ and follow Ref.~\cite{PDG} in defining the ratio $R=\Gamma(\pi\pi)/[\Gamma(\pi\pi)+\Gamma(K\overline{K})]\sim 0.75$, then the \fzKstzOne branching fractions are comparable for the $\fz\ra\pip\pim$ and $\Kp\Km$ channels.

As expected for penguin-dominated channels, the measured \fL values are inconsistent with the na\"ive factorization prediction of $\fL\sim 1$.  The predicted \fL for \frhomKstp is higher than the measured value, though the theory errors are still large.  Including the results from this paper and averaging \babar~\cite{babar_rhoKst2006} and Belle~\cite{BelleRhoKst} \fL measurements for $\rhop\Kstz$, we can order the experimentally measured values of \fL~\cite{babar_rhoKst2006,BelleRhoKst,FergusPRD} as
\beq
\fL(\rhom\Kstp) \lesssim \fL(\rhoz\Kstz) \lesssim \fL(\rhop\Kstz) < \fL(\rhoz\Kstp) ~, \nonumber
\eeq
with the values ranging from $0.38-0.78$.  With the current experimental sensitivities, the three smallest \fL values are consistent with each other at $1\sigma$.  QCDF~\cite{QCDF} predicts the following hierarchy among these \fL values
\beq
\fL(\rhoz\Kstz) < \fL(\rhop\Kstz) < \fL(\rhom\Kstp) < \fL(\rhoz\Kstp) ~, \nonumber
\eeq
which agrees with the experimental finding that $\fL(\rhoz\Kstp)$ is largest.  A more rigorous test of the theoretical hierarchy requires additional experimental input.

For \ffzKstzZero and \ffzKstzTwo, we find
\beqa
\lefteqn{\calB(\fzKstzZero)}  \hspace{3cm}\nonumber \\
\lefteqn{\times\calB(\fz\ra\pi\pi)\times\calB(\KstZero\ra K\pi) }\hspace{2cm} \nonumber \\
  &=& \RfzKstzZero	 ~;	\nonumber \\
\lefteqn{\BfzKstzTwo}  \hspace{3cm}\nonumber \\
    &=& \RfzKstzTwo	~.	\nonumber
\eeqa

For $\rho\KstZero$, we find,
\beqa
\lefteqn{\BrhozKstzZero}  \hspace{3cm}\nonumber \\
		&=& \RrhozKstzZero	 ~; \nonumber \\
\lefteqn{\BrhomKstpZero}  \hspace{3cm}      \nonumber \\
		&=& \RrhomKstpZero 	 ~; \nonumber \\
		&<& \ulrhomKstpZero\times10^{-6} ~. \nonumber
\eeqa

Using the \KstzZeroRes resonant fraction of the LASS \KstzZero result from Ref.~\cite{Latham}, we can calculate the branching fractions for the \KstZeroRes component of our \KstZero channels.  We find
\beqa
\BrhozKstzZeroRes &=& \RrhozKstzZeroRes	 ~; \nonumber \\
\lefteqn{\BfzKstzZeroRes}  \hspace{3.5cm} \nonumber \\
                  &=& \RfzKstzZeroRes	 ~;\nonumber \\
\BrhomKstpZeroRes &=& \RrhomKstpZeroRes	 ~,\nonumber
\eeqa
where the third uncertainty is from the daughter branching fraction $\calB(\KstZeroRes\ra K\pi)=(93\pm 10)\%$~\cite{PDG}.
These results are somewhat lower than the QCDF predictions~\cite{QCDF} but are consistent with QCDF within the uncertainties.  The pQCD predictions have central values of $(0.5-10)\times10^{-6}$ and are, in most cases, inconsistent with our results.

\section{Acknowledgments}
\label{sec:Acknowledgments}

We are grateful for the 
extraordinary contributions of our \pep2\ colleagues in
achieving the excellent luminosity and machine conditions
that have made this work possible.
The success of this project also relies critically on the 
expertise and dedication of the computing organizations that 
support \babar.
The collaborating institutions wish to thank 
SLAC for its support and the kind hospitality extended to them. 
This work is supported by the
US Department of Energy
and National Science Foundation, the
Natural Sciences and Engineering Research Council (Canada),
the Commissariat \`a l'Energie Atomique and
Institut National de Physique Nucl\'eaire et de Physique des Particules
(France), the
Bundesministerium f\"ur Bildung und Forschung and
Deutsche Forschungsgemeinschaft
(Germany), the
Istituto Nazionale di Fisica Nucleare (Italy),
the Foundation for Fundamental Research on Matter (The Netherlands),
the Research Council of Norway, the
Ministry of Education and Science of the Russian Federation, 
Ministerio de Ciencia e Innovaci\'on (Spain), and the
Science and Technology Facilities Council (United Kingdom).
Individuals have received support from 
the Marie-Curie IEF program (European Union) and the A. P. Sloan Foundation (USA).

%-------------------------------------------
% Bibliography

\renewcommand{\baselinestretch}{1}

\end{document}